# A Tactile Tour of the Solar System: Opening a Classic Planetarium Show to Blind and Low Vision Audiences

Briley L. Lewis[1], Ryan Hersey[1], Connor Vancil[1], Skyler Palatnick[1], Maxwell A. Millar-Blanchaer[1], Ila Jade Komasa[2]

[1]University of California, Santa Barbara, Santa Barbara, CA 93106 USA

[2]Santa Barbara Museum of Natural History, Santa Barbara, CA 93105 USA

Abstract: The “tour of the solar system” is a classic offering of many planetaria, but, like much of astronomy, it can be quite inaccessible to blind and low vision audiences. This work presents 3D-printed tactile models for five new solar system objects—Mercury, Venus, Saturn, Neptune, and Pluto—to make this common outreach topic open for new audiences to experience through the sense of touch. When combined with previously published models, there now exists a nearly complete solar system model set for this intended use. We describe the details of the models’ design, report results from usability testing, and provide recommendations for best uses of these resources.

## 1. Introduction

For over 50 years, the planetarium has been a cornerstone of astronomy education, particularly useful for inspiring students’ interest in the field and providing opportunities for interactive demonstrations (Burnette Jr, 1976; Petrie, 2013; G. Reed, 1970; G. F. Reed, 1970; Ridky, 1974; Rusk, 2003). However, astronomy—especially in the traditional planetarium—is a highly visual expe-rience, making it inaccessible to blind and low vision (BLV) audiences. BLV students are also often at disadvantages in the STEM classroom, due to a lack of accessible graphical information (Beck-Winchatz & Riccobono, 2008).

A number of efforts have tried to bridge this gap, with data sonification (Harrison, Zanella, & Girdhar, 2024) or tactile materials: e.g. *Varano and Di Giacomo (2024)*’s multisensory planetarium, *Yajima, Toshimitsu, and Tetsuya (2014)*’s tactile star wheel, *Cacace and Pereyra (2022)*’s haptic feedback, *Suzuki, Ushida, and Chen (2024b)*’s touchable virtual reality planetarium, *Suzuki, Ushida, & Chen (2024a)*’s vibrating planisphere, and *Trotta et al. (2020)*’s olfactory planetarium.

Creating these multi-sensory resources can often be difficult or resource-intensive, requiring specialized thermoform printers or other technology. Many outreach practitioners are also not familiar with the needs of the BLV commu-nity and/or don’t know where to start on such an endeavor, so putting together easy-to-use resources can be impactful to improve uptake of these accessible activities (Ortiz-Gil et al., n.d.). There are multiple options for lower-cost tactile outreach that have been considered (e.g. foam, pipe cleaners, beads, yarn) (Grice, 2022), and 3D-printing—especially with the proliferation of cam-pus makerspaces—is emerging as a particularly useful technology (Christian et al., 2025; Madura et al., 2025; Wonjin et al., 2016). Previous efforts have cre-ated tactile adaptations of many *Hubble Space Telescope* images (Christian, Nota, Greenfield, Grice, & Shaheen, 2015; Christian, Nota, Grice, Madura, & Hurd, 2025; Grice, Christian, Nota, & Greenfield, 2015) and *Chandra X-ray Telescope* data (Arcand et al., 2019; Arcand, Price, & Watzke, 2020), as well as a tool (astro3d) to easily convert astronomical images of stars and galaxies to 3D models using a user-friendly GUI[1].

This work aims to create models for a different subset of astronomical imagery: the solar system planets as seen by various NASA missions. So-called "tours of the solar system" are a commonly-used lesson in introductory astron-omy courses, planetarium shows, and other outreach events; when discussing with our local planetarium, we determined that solar system tactile models were likely to have the largest impact on accessibility of their programming, as the solar system shows were the most often offered at that institution. In the planetarium, the so-lar system can also be used as a starting point for teaching the scales of the universe (Yu, Sahami, & Dove, 2017) and illustrating fundamental concepts related to the Moon and its orbit around Earth (Yu, Sahami, Denn, Sahami, & Sessions, 2016)—two other fundamental lessons common to introductory astronomy.

In this work, we present designs for five new 3D-printed solar system tac-tile models: Mercury, Venus, Saturn, Neptune, and Pluto. In combination with existing 3D-printing resources, there should now be a nearly complete set of solar system models to create an accessible "tour of the solar system" for blind and low vision audiences. This work describes the design choices in the mod-els, and presents results of informal usability testing of the new designs. The models are also made publicly available as .stl files.

## 2. Model Design

Models for Earth, Earth's Moon, Mars, and Jupiter were created by Carol Christian (STScI) and the files are publicly available online (Christian, n.d.); these models have been previously tested with blind and low vision audiences for usability. Models for Mercury, Venus, Saturn, Neptune, and Pluto are all newly presented in this work, and were created using Blender with a combination of real data, artist's renditions, and manually-added textures, similar to the process used in Carol Christian's existing designs. In creating tactile models, it is advised to consider what message you want to communicate, and tailor the graphics accordingly, smoothing out unnecessary detail and exaggerating fea-tures of interest (Ortiz-Gil et al., 2019). Note that Uranus was not included in this set, as the data on its surface is more limited and its surface features are quite subtle; in lessons, we recommend describing Uranus when participants are experiencing Neptune or using other NASA-created resources such as the Uranus tactile panel from Space Telescope Science Institute[2]. Multiple iterations of test prints were created and experienced by members of the development team; this helped us to find a balance between features that were of an appropriate size and exaggerated enough to be noticeable, but not so exaggerated as to be confusing or unpleasant to the touch. The overall size of the model can be adjusted by the end user, scaling the .stl file to the desired size in any 3D-printer slicing software of the user's choice.

The Mercury tactile model was created by wrapping the *MESSENGER* topographic map around a sphere, then smoothing the map and exaggerating the elevation of features to make them easier to feel (Hunter et al., 2018; Preusker et al., 2011). The goal of the Mercury tactile model is to illustrate how barren, rocky, and thoroughly cratered the planet

[1] https://github.com/spacetelescope/astro3d
[2] https://outerspace.stsci.edu/spaces/STTI/overview

is, somewhat similar to Earth's moon.

Figure: IMG_3181.jpeg
Caption: Tactile model of Mercury.

The Venus tactile model was created from two spheres split in half and spliced together; one half is the exaggerated surface topography from *Magellan* topographic maps of the surface (Ford, 1993), and the other shows the visible outer cloud layer from a NASA artist's rendition based on *Mariner 10* imagery. The cloudy half of the tactile model also has a manual texture around the center to help convey the idea of the fluffy, somewhat banded cloud structure, similar to the aforementioned Jupiter models. Using this dual-sided design, an audience member would be able to experience both the cloud layer (i.e. what Venus looks like to a spacecraft flying by) and the rocky surface shaped by vol-canism. This would provide a presenter with an opportunity to discuss remote sensing techniques (e.g. radar used to probe the surface), as well as the thick atmosphere of Venus and its runaway greenhouse effect.
Figure: IMG_3185.jpeg
Caption: Tactile model of Venus, illustrating the split design described here.

The Saturn tactile model was the most "idealized" of the newly created designs. The subtle surface texture is created from Björn Jónsson's plane-tary maps, which use *Cassini* and *Voyager 2* data (Jonsson, n.d.; Lebreton & Matson, 1992; Smith et al., 1982). The most noticeable feature on the tactile Saturn—other than its rings—is the polar hexagon (Allison, Godfrey, & Beebe, 1990; Yadav & Bloxham, 2020), which is additionally exaggerated to make it no-ticeable to a user. The rings are also highly simplified and not to scale, although they do include a gap to illustrate the idea of the Cassini division (Cuzzi, Filac-chione, & Marouf, 2018; Goldreich & Tremaine, 1978). The goal for the Saturn model is to provide a rough sense of the smooth globe surrounded by highly prominent rings with a noticeable gap, and the inclusion of the polar hexagon provides an opportunity to discuss this strange, intriguing feature on the gas giant.

Figure: IMG_3202.jpeg
Caption: Tactile model of Saturn. The split coloring was due to a change in filament halfway through the print.

The Neptune tactile model was, like Saturn, created from Björn Jónsson's planetary maps; these maps for Neptune were created from *Voyager 2* data (Smith et al., 1989), as well as some artistic rendering due to the lack of full global coverage from the spacecraft's flyby (Jonsson, n.d.). Two swirls (one large and one small) were added to represent Neptune's Great Dark Spot and Small Dark Spot as seen by *Voyager 2*; these are the most prominent features on the tactile model, and provide an opportunity for a presenter to discuss planetary storms and how they have changed since this snapshot by *Voyager* (Hsu, Wong, & Simon, 2019; Polvani, Wisdom, DeJong, & Ingersoll, 1990).

Figure: IMG_3203.jpeg
Caption: Tactile model of Neptune, with the tactile representation of the Great Dark Spot.

The Pluto tactile model was created with the digital elevation map (Schenk et al.,

2016) from the *New Horizons* mission (Stern, 2008), with a manually-added exaggeration of the distinctive “heart” shape on Pluto’s surface (Sputnik Planitia and Tombaugh Regio) (Moore et al., 2016; Stern et al., 2015). Some of Pluto’s most notable craters (e.g. Burney crater) are further enhanced to make them stand out to a user. The Pluto tactile model notably does *not* have an artistic rendering or fictional data to fill in the portion of the globe not imaged in the flyby, and some portions of the map are notably lower resolution (as the probe was further from the dwarf planet while taking those images); this provides an opportunity for the presenter to explain the nature and limitations of the flyby type of mission.

Figure: IMG_3208.jpeg
Caption: Tactile model of Pluto, with the manually added “heart” of Sputnik Planitia.

These designs were 3D-printed using a Bambu Labs P1S printer with PLA filament, using variable layer heights to increase the detail achieved in the prints. Braille labels were added to some of the models for added clarity, including labels for “North” and the planet’s name. Note that it is *not* recommended to 3D-print Braille (Reinders, Ananthanarayan, Butler, & Marriott, 2023); accord-ingly, these labels were created with an analog Reizen RL-350 Braille Labeler. Notably, Mercury did *not* have any labels attached, as the planet’s texture was too rough for the labels to stick without extra glue. Other models had the fol-lowing additional notable features labeled: clouds (Venus), rings (Saturn), great dark spot (Neptune), Sputnik Planitia (Pluto), no data (Pluto).

Figure: IMG_3178.jpeg
Caption: The 3D-printed solar system, including both models from C. Christian (Earth, Moon, Mars, Jupiter) and this work (Mercury, Venus, Saturn, Neptune, Pluto).

## 3. Usability Testing

***Methods:***

To gather feedback on these models, we conducted informal “lightly-moderated” usability testing similar to that of *Christian et al. (2025)*. The goal was to under-stand how easy it is for participants to interpret the models and what aspects of the experience they enjoy, are confused by, dislike, etc. to further improve the designs. At an informal outreach event, we shared the models with five blind and low vision attendees; three of these participants were fluent with Braille, one was learning, and one was unfamiliar. The one participant unfamiliar with Braille also identified as sighted, but with low vision. These were adults, and one had a strong interest in astronomy with some prior knowledge.

In this outreach event, we introduced participants to the models in very small groups (one-on-one or in pairs), providing substantial interaction between volunteer educators and participants. Participants were handed a model ori-ented with North up, and told which planet it is. They were then asked what shapes, textures, and features they noticed, and what their impression of the planet is. These observations formed a starting point for a conversation about each planet. For example, when participants noted that Mercury was incredibly rough and rugged, we then talked about cratering and its lack of atmosphere to protect it from impacts. If the participant was a Braille reader, we pointed them towards Braille labels for certain features as well. During this exploration, we took notes on the behavior, expressions, and verbal feedback of the participants.

At the end of the outreach session, we asked for general feedback using the following questions:

- How can we improve these models? What did you like or dislike?
- What other celestial objects would you like to have 3D models of?
- Has working with these models had any effect on your understanding of and/or interest in astronomy?

***Results:***

Participants noted that Mercury was very bumpy, which one person correctly interpreted as filled with craters. When presented with a model of the Moon for comparison, they noticed the similarities between these barren, cratered worlds.

For Venus, they noticed the cloud bands and the various textures used to differentiate them, with one participant specifically noting the linearity of the clouds along the latitudinal direction (i.e. that they do appear in bands or stripes, interpreted with the correct orientation). It is worth noting that some participants initially thought the textures were actual features, so it is impor-tant for the facilitator to guide the participants on what they are feeling. On the surface side, they noticed that it was again very crater-heavy and rough. One participant commented that they didn't find much identifiable on the surface at first touch, but taking more time they found more and more details including mountains.

Saturn's most notable feature was, unsurprisingly, its prominent rings—including the gap between them. Participants also noticed how smooth it was, and that it was somewhat bulged at the equator. They also noted the hexagonal storm at the North pole after some prompting to explore there; for one participant, it was difficult at first to tell that this storm was a hexagon and not just a smooth circle.

Participants noticed that Neptune was generally smooth and relatively featureless, but the large bumps of clouds stood out to them. The Great Dark Spot and Small Dark Spot were immediately noticeable. When directed to compare Neptune with another giant planet like Saturn, and contrast it with a rocky planet like Mercury, participants were clearly able to grasp the distinction between the two planet types via the tactile models.

Lastly, Pluto's most notable features were the heart and some of its larger craters. Participants were particularly excited to learn about Pluto, calling the model "awesome" and expressing enthusiasm that this world was included in addition to the major planets. They were surprised at how bumpy the surface was, and noticed the varying levels of texture: very bumpy at the point of *New Horizons*' closest approach, slightly smoother around the back, and completely smooth in the area labeled "no data." This presented a clear opportunity to discuss the flyby and its limitations for data collection, and to emphasize the outstanding mystery of what else is on the side of Pluto we haven't seen, which seemed to inspire a great deal of curiosity and wonder.

Nearly all first reactions to the models included noises of surprise or curiosity like "ooh", and participants frequently remarked on how cool, fascinating, interesting, etc. they found the models. Participants were highly engaged and actively exploring the models for the entire duration of the event, and even began to interact with each other to show their friends the different features and things they learned about on a given model. When asked about what they liked, participants stated how much fun they had and that they enjoyed the whole experience and all the models they got to touch. Multiple participants also remarked that the size of the models (approximately 6 inches across) was excellent for being able to hold with two hands and get an idea of the planets' shapes. They did note that Saturn should be printed larger; our first model had Saturn printed with a 6 inch diameter *including* the rings, and they stated a preference for the planet body to be 6 inches itself to match the other globes. Making Saturn larger would also likely ameliorate the issue of determining the hexagonal shape of the polar storm. For the one sighted participant, they men-tioned that the colors (which were quite bright) were a bit odd at times, suggesting perhaps finding more subtly colored 3D-printing filaments.

The Braille labeling had mixed reactions. Although some found them helpful (especially for exploration without as much guidance), some participants noted that fewer people are reading Braille nowadays due to improvements in technology, and at first it may be hard to distinguish between the Braille and smaller surface features. This led us to make one notable change to the models—instead of adding a Braille label to mark North, we have now added a small "pointer" nodule to indicate it in the prints themselves.

When asked what other celestial objects they would like to have 3D printed, participants said *anything*—they emphasized how they would be eager to try any other new experiences of astronomical data in tactile form, and specifi-cally mentioned the recently-discovered "interstellar asteroids" like 1I/2017 U1 'Oumuamua. Participants also stated that these models taught them something new, and eagerly encouraged the volunteers to come back with new models to try out whenever they're ready.

## 4. Guidelines for Model Usage and Example Lessons

Here we present recommendations for the use of these models in the form of a "how-to" guide for potential users.

***Printing the Models:***

The models are .stl files ready-to-go for slicing in any 3D-printing slicer soft-ware of the user's choice. The user can re-scale the planets to their size of choice; we set the diameter of each to 150 mm (5.9 inches). The amount of filament used is approximately 500 grams per globe, which for our filament (BambuLabs PLA) cost approximately $12, and each took around 18-30 hours to print.

***Best Practices for Facilitation:***

Many excellent resources exist for best practices for BLV outreach more generally, e.g. a guide from the Astronomical Society of the Pacific[3]. In the spirit of lowering barriers to entry in this outreach work, we summarize some of the most relevant points here and add guidance specific to our models:

- Include accessibility information on your program's website, e.g. avail-ability of tactile models, other accommodations offered, who to contact with questions.
- Consider the physical accessibility of your space, and if needed, arrange for volunteers to act as guides. *Always ask first if assistance is desired, especially*

[3] https://nightsky.jpl.nasa.gov/media/documents/resources/BestPracticesBLVVisitorsMuseums2023.pdf

*before touching someone.*

- Make sure to verbally introduce and describe yourself to visitors, and to let them know when you are leaving.
- Use visitors' names when addressing them so they know you are speak-ing to them.
- Give specific directions, such as "the model is two inches in front of your hands" instead of "the model is right there."
- Hand the models to participants with North up, and tell them that you have done so.
- If the participant is a Braille reader, tell them where to expect the la-bels. If they have trouble finding the labels, ask for permission then guide their hands to the relevant label. Note that Braille readers often use both hands, not just their dominant hand.
- The ideal ratio of models to people is 1:1. Using these models takes time to explore; you can't briefly flash these models like a photo on a slide, and only one person at a time can effectively experience a single model.
- Similarly, this type of exploration will likely require additional time compared to an image-based planetarium show. Regularly scheduling a longer showtime to encourage use of these models may be beneficial, normalising their use for not only blind and low vision individuals, but any participants who may be interested..
- Similarly, it is ideal to have a facilitator guiding a participant in their exploration of the models, at least one facilitator per 1-3 participants. It is not sufficient to hand someone a model without context during an otherwise visually-based planetarium show and assume they will figure it out, and can in fact cause harm as a participant may feel neglected and confused.
- It is key to make connections with your local BLV groups, and to under-stand what the disabled population in your community actually wants and needs.
- Program coordinators should ensure that staff and volunteers are made aware of the tactile models available, and are therefore able to recom-mend them to visitors who may require accommodations. Similarly, program coordinators should brief staff and volunteers on when/how to use the models and best practices for doing so. Coordinators may also wish to maintain a "cheat sheet" or other document specific to their institution to accompany the materials with both best practices and important features to point out to users.

***Intended Learning Goals and Notable Features for Each Model:***

These models are designed to give an impression of each planet, similar to their classic photos shown in planetarium shows; these intended impressions are described below, along with some ideas of key points or notable features for a facilitator to include. Facilitators should also be prepared with general so-lar system facts, as they would for any solar system event or show.

**Mercury:** Participants will note that Mercury has an incredibly, rough cratered surface, as it has no atmosphere to protect it due to its proximity to the Sun.

**Venus:** Participants will note that Venus's surface and its cloud deck are quite different; they will learn that the clouds are so thick, they're all we see unless we use radar to "see" the surface. Its surface is rocky, lifeless, and full of vol-canic activity.

**Saturn:** Participants will note that Saturn is much smoother than the rocky planets, because it is instead a *gas giant* with no solid surface. Participants will also note the prominent rings. A facilitator may also point out the famous hexagonal storm at the North, or the Cassini division in the rings.

**Neptune:** Participants will note that Neptune is also quite smooth like Saturn, a gas (or ice) giant as opposed to the rocky inner worlds like our own. They will notice its clouds, and learn about persistent storms on the giant planets like the Great Dark Spot and Small Dark Spot.

**Pluto:** Participants will likely notice the prominent heart, which a facilitator can explain; half of the heart is Sputnik Planitia, a crater filled with nitrogen ice to create a large, flat plain, because Pluto is so far from the Sun and so cold that it's made of mostly ice. Participants will notice the difference in the "resolution" of the textures. Some regions are smoother, and the Southern region is completely smooth; the facilitator will explain the nature of the data from the *New Horizons* flyby, where the rough texture is actually where we have more detail, the smoother back is where we have lower-resolution data, and the smooth bottom is where we have no data at all.

## 5. Discussion and Conclusions

In this work, we created a new set of 3D-printed tactile models of solar system objects: Mercury, Venus, Saturn, Neptune, and Pluto. Together with previously-developed designs from NASA/Space Telescope Science Institute, there now exists a full solar system of 3D models for blind and low vision outreach; this is particularly useful as many planetaria host shows with the theme "a tour of the solar system," so these resources can make that common programming more accessible. Our new designs were informally tested with a group of blind and low vision participants, and were quite well received. Participants expressed excitement at the opportunity to explore the models, and were able to extract the intended lessons on solar system planets' characteristics from the models with appropriate guidance.

From this experience, we learned multiple lessons for future improvement, and for best usage of the models. Models should be printed with colors that are either representative of the planets or are more neutral, to avoid confusing sighted low vision participants. Braille labeling can be helpful (especially when one-on-one facilitation is not possible), but need to be placed with care on models that have a lot of rough small textures similar to Braille dots; alternatively, the labels can be pointed out clearly to participants by the facilitators, or other design choices could be explored, such as adding a clear border to the label. It is also helpful to have a facilitator to work closely with participants as they explore the models, which echos prior findings for blind and low vision tactile model outreach from Madura et al. (2025) that tactile models need to be introduced with appropriate context to be effective. Similarly, there needs to be at least one model per person, as it takes time to explore these 3D-printed worlds (Usuda-Sato, Nakayama, Fujiwara, & Usuda, 2019).

Participants made it clear that they were strongly enthusiastic about en-gaging with the models, illustrating the clear need for accessible outreach for blind and low vision audiences. To make more outreach like this possible and to lower the barrier to entry for other outreach practitioners, we have made these 3D-printed designs publicly accessible, accompanied by best practices and guidance in this paper. With just access to a 3D printer, other outreach ef-forts can create similar materials and bring the solar system to their local blind and low vision communities, opening participation in astronomy to a traditionally underserved community.

## 6. Notes

***Data Sources:*** The model prints developed in this work are available on Zenodo at 10.5281/zenodo.18842652, and the previously developed models from C. Christian are available through the Space Telescope Science Institute As-tro3D website: https://www.stsci.edu/carolc/3dastronomy/gallery4.html. Cylin-drical elevation maps for the surfaces of Mercury, Venus, and Pluto are avail-able through USGS Astrogeology: https://www.usgs.gov/centers/astrogeology-science-center/maps.
Imagery of Venus, Saturn, and Neptune is available via NASA, although some maps used herein were from a re-processing of NASA resources by Björn Jónsson.

***Funding:*** National Science Foundation Astronomy & Astrophysics Postdoctoral Fellowship Award No. 2401654

***Ethics:*** The informal usability testing in this work was not required to have IRB certification from UCSB; documentation attesting this can be provided upon request.

## 7. Acknowledgements

The authors would like to thank the Braille Institute of Santa Barbara (particularly Mary Rocha for coordinating efforts, and all our outreach event participants) and the Santa Barbara Museum of Natural History for their collaboration. Thank you to Carol Christian, Tom Madura, Cherilynn Morrow, and other members of the accessible astronomy outreach community for sharing their wisdom on creating tactile materials and helping to build a stronger community of practice in this area.

This material is based upon work supported by the National Science Foundation Astronomy & Astrophysics Postdoctoral Fellowship Award No. 2401654. Any opinions, findings, and conclusions or recommendations expressed in this material are those of the authors(s) and do not necessarily reflect the views of the National Science Foundation.